\documentclass{iucrjournals}

\usepackage{siunitx}

\title{A compact fibre-coupled active beamstop for fast time-resolved monitoring of synchrotron X-ray beams}

\author[a]{Friedjof Tellkamp\IUCrCemaillink{friedjof.tellkamp@mpsd.mpg.de}\IUCrOrcidlink{0000-0002-5670-3544}\IUCrAufn{These authors contributed equally.}}%
\author[b]{David von Stetten\IUCrCemaillink{dvonstetten@embl-hamburg.de}\IUCrOrcidlink{0000-0001-7906-9788}\IUCrAufn[1]{}}%
\author[a]{Martin Kollewe}%
\affil[a]{Scientific Support Unit Machine Physics, Max Planck Institute for the Structure and Dynamics of Matter, Luruper Chaussee 149, 22761 Hamburg, Germany}
\affil[b]{European Molecular Biology Laboratory (EMBL), Hamburg Unit c/o DESY, Notkestraße 85, 22607 Hamburg, Germany}

\begin{document} 
\maketitle 

\begin{synopsis}
A compact fibre-coupled active beamstop with a maximum tip width of \SI{1}{mm} combines direct-beam interception with time-resolved monitoring. The device is fast enough to distinguish individual bunch signals separated by \SI{192}{ns} at PETRA III and provides an in-situ timing signal for synchronising an experimental sequence with a mechanical X-ray chopper during diffraction data acquisition.
\end{synopsis}

\begin{abstract}
We have developed a compact fibre-coupled active beamstop for simultaneous interception and time-resolved monitoring of the direct X-ray beam in synchrotron diffraction experiments. The complete beamstop tip has a maximum transverse dimension of \SI{1}{mm} and incorporates a \SI{700}{\micro m}-wide Ce:YAG scintillator. The scintillator converts the intercepted X-rays into visible light, which is transported through a multimode optical fibre to a silicon photodiode at the bottom of the blade that is connected to a dedicated fast amplifier electronics. Tests with a pulsed laser yielded a system decay time of \SI{69.4}{ns}, consistent with reported luminescence decay times for Ce:YAG. Measurements at the P14.EH2 end station of PETRA III demonstrated the detection of individual bunch signals separated by \SI{192}{ns} in the 40-bunch operating mode. The active beamstop also provided an in-situ timing signal for synchronising an experimental sequence with the approximately \SI{190}{\micro s} transmission window of a mechanical X-ray chopper. The device therefore combines a compact obstruction of the direct beam with fast monitoring of X-ray timing and flux during diffraction data acquisition.
\end{abstract}


\section{Introduction}

In synchrotron diffraction and scattering experiments, the direct X-ray beam is generally several orders of magnitude more intense than the diffracted or scattered signal. A beamstop is therefore placed downstream of the sample to protect the direct beam from reaching the sensitive detector. In experiments with short sample-to-detector distances, including typical macromolecular crystallography geometries, the beamstop must intercept the direct beam while obscuring as little as possible of the low-angle diffraction region. Its transverse dimensions and supporting structure consequently impose important constraints on the accessible scattering angles.

Being able to track X-ray intensity and timing during the actual data acquisition is also valuable. Conventional diagnostics, including ionization chambers, photodiodes and beam-position monitors, are commonly located upstream of the sample. Depending on their positions, these monitors may not detect changes introduced farther downstream by shutters, choppers, apertures, collimators, sample-environment components or the sample itself. For example, slight lateral vibrations of the X-ray beam might not be detected by upstream devices as long as the overall intensity does not change, but a small beam-defining aperture close to the sample position will translate such vibrations into a modulation of the X-ray flux interacting with the sample. Measuring the direct beam downstream of the interaction region can therefore provide complementary information on the X-ray flux during an exposure for every recorded image. In addition to monitoring intensity, a sufficiently fast downstream signal can be used to characterise beam timing and to synchronise experiments employing pulsed beam delivery by fast shutters or mechanical choppers.

A range of active beamstop concepts has previously been developed. These include a PIN diode embedded in a tungsten--epoxy absorber \cite{Ellis2003}, photoelectron collection from an electrically isolated beamstop \cite{Pan2014}, detection of radiation back-scattered from a cavity in the absorber \cite{Blanchet2015}, and a single-crystal CVD diamond detector mounted within a tungsten-carbide absorber \cite{Desjardins2021}. These devices were developed principally for relative or absolute flux measurement as well as beam alignment, which defined the required dynamic range and radiation tolerance. Beamstop-mounted photodiodes have also been used for pulse-resolved normalisation of transmitted intensity at an X-ray free-electron laser \cite{Muller2022}.

Spatial separation of the X-ray conversion element from the photosensor has also been demonstrated. Englich \textit{et al}.\ used a CdWO$_4$ scintillator and a flexible light guide to transport the generated light to a remote photodiode, providing simultaneous beam interception and flux monitoring for microfocus beams \cite{Englich2011}. A similar fibre-coupled implementation employing a Ce:GAGG scintillator in a tungsten absorber is commercially available with a nominal beamstop diameter of approximately \SI{1.25}{mm} \cite{MiTeGenSentinel}. These implementations are designed primarily for beam intensity and dose monitoring. By contrast, the present work emphasizes temporal response on the scale of the synchrotron bunch spacing while retaining a compact beamstop tip.

Here, we present a fibre-coupled active beamstop with a maximum transverse tip dimension of \SI{1}{mm}. An embedded \SI{700}{\micro m}-wide cerium-doped yttrium aluminium garnet (Ce:YAG) scintillator converts the intercepted X-rays into visible light, which is transported through a multimode optical fibre sitting in a groove of the beamstop blade to a silicon photodiode and fast amplifier electronics positioned outside the diffraction cone. The use of Ce:YAG and a dedicated fast readout distinguishes the present implementation from fibre-coupled systems based on slower intensity readout and enables direct observation of rapid changes in X-ray flux. We describe the mechanical, optical and electronic implementation, characterise its electronic noise and temporal response, and evaluate the device at the P14.EH2 (T-REXX) end station of PETRA III. The device is able to resolve X-ray pulses from individual electron bunches separated by \SI{192}{ns} in the PETRA III 40-bunch operating mode and provides an in-situ signal for synchronizing an experimental sequence with the transmission window of a mechanical X-ray chopper during diffraction data acquisition.


\begin{figure}[th] %
\begin{center}
\includegraphics[width=0.5\textwidth]{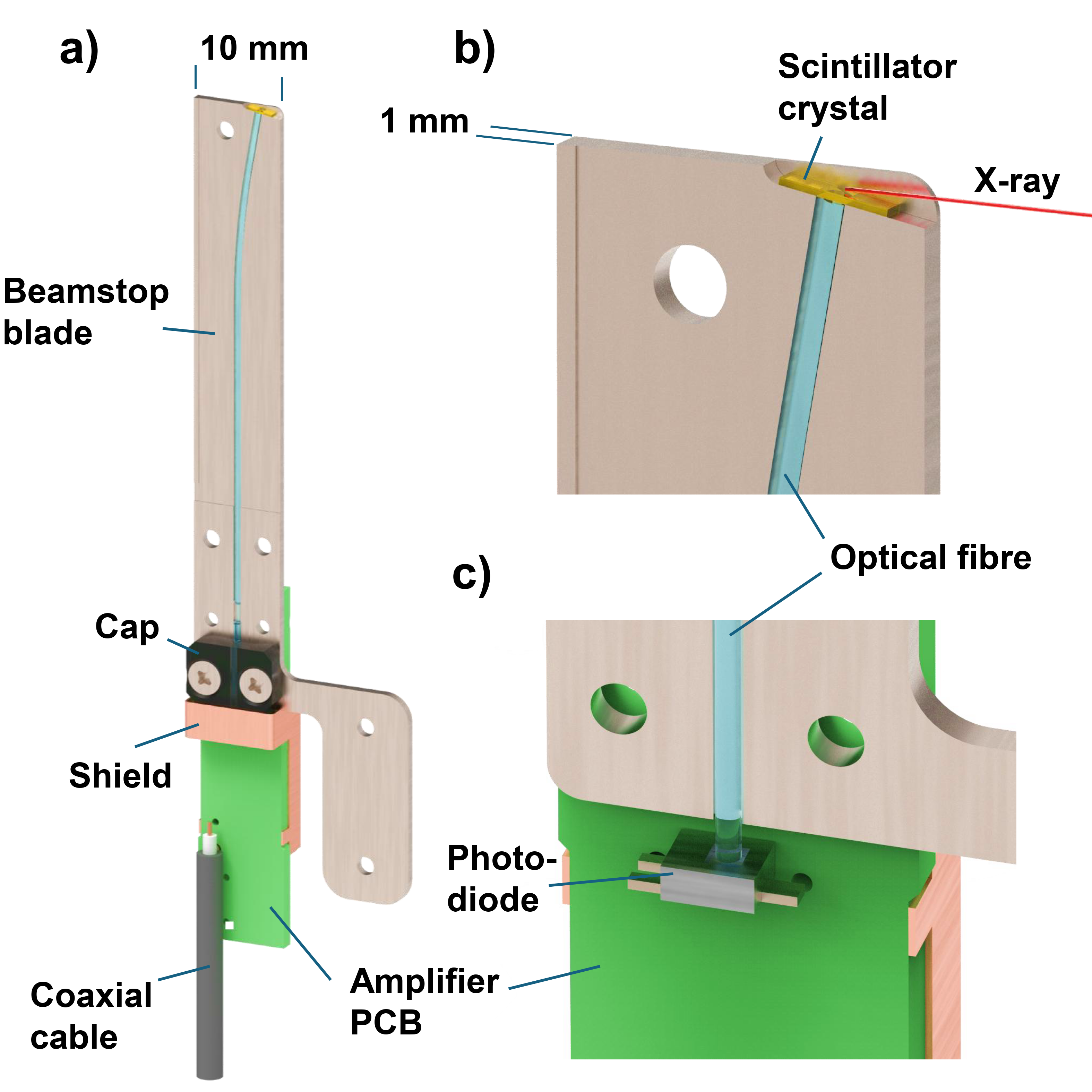} 
\end{center}
\caption{Rendered views of the fibre-coupled active beamstop assembly.
(a)~Overview of the complete assembly.
(b)~Detailed view of the beamstop tip showing the Ce:YAG scintillator, fibre-coupling geometry and direction of the incident X-ray beam.
(c)~Detailed view of the photodiode and amplifier assembly with the protective cap and electromagnetic shield removed.}
\label{fig:CAD_render}
\end{figure}

\section{Active beamstop design}

The principal geometrical requirement was to minimise the transverse dimension of the beamstop in order to preserve the low-angle diffraction region. A target width of approximately \SI{1}{mm} or less precluded placement of a photodiode and readout electronics directly at the beamstop tip. The intercepted X-rays are therefore converted into visible light by a Ce:YAG scintillator, and the scintillation light is transported through an optical fibre to a silicon photodiode positioned outside the diffraction cone. The photodiode and amplifier electronics are mounted on the lower part of the beamstop blade. An overview of the resulting assembly is shown in Fig.~\ref{fig:CAD_render}.

\subsection{Scintillator geometry and fibre coupling}

Ce:YAG was selected as the scintillator because of its mechanical robustness, commercial availability, visible emission and relatively short luminescence decay time. Reported room-temperature values range from approximately \SI{63}{ns} to \SI{80}{ns}, with values near \SI{70}{ns} commonly observed \cite{Blasse1967,Blasse1994,Bachmann2009}. At an X-ray energy of \SI{12.7}{keV}, the attenuation length in YAG is approximately \SI{100}{\micro m} \cite{Henke1993}. The X-ray attenuation provided by the scintillator therefore depends on both its physical thickness and its orientation relative to the beam. Behind the Ce:YAG scintillator approximately \SI{5}{mm} of steel ensure at least \SI{99.999}{\percent} absorption up to \SI{43}{keV} in the beamstop tip.

A double-side-polished Ce:YAG crystal with a nominal Ce doping concentration of \SI{0.2}{\percent}, as specified by the manufacturer, was obtained from EPIC Crystal (China). The crystal was subsequently ground in-house to final dimensions of \SI{700}{\micro m} $\times$ \SI{200}{\micro m} $\times$ \SI{2}{mm} (width $\times$ thickness $\times$ length). The width was selected to maintain a small obstruction at the direct-beam position, while the length provided sufficient surface area for mechanically stable attachment to the beamstop blade. The crystal was mounted at an angle of approximately \SI{10}{\degree} between the crystal surface and the horizontally incident X-ray beam. This orientation produced an effective X-ray path length in YAG of approximately \SI{1.1}{mm} and shifted the primary X-ray absorption region towards the side of the crystal opposite the fibre-coupling surface. Figure~\ref{fig:CAD_render}b shows the crystal orientation and the direction of the incident beam.

Ce:YAG emits over a broad spectral range from approximately \SI{500}{nm} to \SI{650}{nm}, with an emission maximum near \SI{550}{nm} \cite{Bachmann2009}. The responsivity of the selected silicon photodiode at this wavelength is approximately \SI{0.25}{A/W}. Using a representative Ce:YAG scintillation yield of the order of \SI{20e3}{photons/MeV} \cite{Moszynski1994}, an incident photon flux of \SI{1e12}{ph/s} at an X-ray energy of \SI{12.7}{keV}, and an assumed overall optical collection efficiency of \SI{0.1}{\percent}, the expected photodiode current is of the order of \SI{20}{nA}. This order-of-magnitude estimate was used to determine the required amplifier gain; the light yield and optical collection efficiency were not measured for the present device.

No imaging or focusing optics were incorporated because of the restricted space at the beamstop tip. Instead, a bare multimode optical fibre, cleaved on both ends, was placed in direct contact with an optically polished planar surface of the scintillator, without the use of any index-matching gel. The fibre was an FG600UEA multimode fibre (Thorlabs) with a core diameter of \SI{600}{\micro m} and a numerical aperture of 0.22.

The beamstop blade was machined with an inclined pocket for positioning the scintillator and a longitudinal groove for the optical fibre. The scintillator was fixed to the beamstop tip using epoxy adhesive, and the fibre was retained using adhesive tape. The upper surface of the beamstop body was partially covered with \SI{50}{\micro m}-thick copper tape to reduce the scattering of stray radiation towards the detector. The maximum transverse dimension of the complete assembly at the direct-beam position is \SI{1}{mm}. A detailed drawing of the beamstop blade is given in Fig.~S2. 

\subsection{Photodiode and amplifier electronics}

At the opposite end of the optical fibre, the scintillation light is detected using an SFH~2400 silicon photodiode. The photodiode has an active area of \SI{1}{mm} $\times$ \SI{1}{mm}, which is sufficiently large to receive the light emerging from the \SI{600}{\micro m}-core fibre while retaining a sufficiently small junction capacitance for fast operation. The photodiode is operated with a reverse-bias voltage of \SI{5}{V}.

No imaging optics are used between the fibre and the photodiode. Instead, the fibre end is placed in direct contact with the active area of the photodiode [Fig.~\ref{fig:CAD_render}c]. The photodiode is mounted on a compact printed circuit board (PCB) that also accommodates the amplifier electronics. The PCB measures \SI{10}{mm} $\times$ \SI{40}{mm} and is attached to the lower part of the beamstop blade, outside the diffraction cone. A light-tight, 3D-printed cap encloses the fibre--photodiode interface to reduce sensitivity to ambient light.

The amplifier comprises three LMH6624 operational-amplifier stages arranged as one transimpedance stage followed by two voltage-amplification stages. The first stage uses a feedback resistance of \SI{10}{k\ohm} and a feedback capacitance of \SI{3.3}{pF}, giving a nominal low-frequency transimpedance of \SI{10}{V/mA}. The feedback-network time constant, $R_{\mathrm{f}}C_{\mathrm{f}}$, is approximately \SI{33}{ns}, corresponding to a first-order pole frequency of approximately \SI{4.8}{MHz}. This value characterises the feedback network rather than the bandwidth of the complete readout chain. Each of the two subsequent voltage-amplification stages have a gain of 10, resulting in an overall nominal low-frequency transimpedance of \SI{1}{V/\micro A}. The PCB is designed to drive a \SI{50}{\ohm} load and includes provision for optional output-offset compensation.

The amplifier is powered through twisted-triplet wiring from an LDS12B bipolar power supply (Thorlabs). The externally supplied \SI{\pm 12}{V} rails are regulated to \SI{\pm 5}{V} before being supplied to the amplifier circuit. The output signal is transmitted through an RG-174/U coaxial pigtail cable soldered directly to the PCB.

To reduce pickup of external electromagnetic interference, the photodiode and input-amplifier region was enclosed on both sides by copper sheet-metal shields. The effect of this shielding on the measured electronic noise is described in Section~\ref{sec:res_dis}.

\section{Experimental characterisation}\label{sec:exp_char}

\subsection{Electronic-noise measurements}

The electronic noise of the active beamstop was measured using a Teledyne LeCroy WaveSurfer 3404z oscilloscope with a \SI{50}{\ohm} input termination. During the measurements, the scintillator was not illuminated and the fibre--photodiode interface was enclosed by the light-tight protective cap. Measurements were performed with and without the copper shielding around the photodiode and amplifier region while all other experimental conditions were kept unchanged.

For each shielding configuration, 32 waveforms were acquired at a sampling rate of \SI{500}{kS/s}, each covering an acquisition interval of \SI{200}{ms}. The noise was quantified from the standard deviation and peak-to-peak amplitude of the output voltage using the oscilloscope's built-in statistical measurement functions. No additional digital filtering was applied.

\subsection{Pulsed-laser measurements}\label{subsec:pulsed_laser_meas}

The temporal response of the active beamstop was characterised by direct optical excitation of the Ce:YAG scintillator using a Thorlabs NPL45B pulsed laser with an emission wavelength of \SI{450}{nm}. The specified optical pulse width of \SI{39}{ns} was shorter than the expected Ce:YAG luminescence decay time of approximately \SI{70}{ns}. The laser was operated at a repetition rate of \SI{1}{MHz}, corresponding to a pulse separation of \SI{1}{\micro s}. The unfocused laser beam illuminated the scintillator from the same direction as the X-ray beam in the intended operating geometry. The active beamstop output was recorded using a Keysight DSOX3034T oscilloscope with a \SI{50}{\ohm} input termination, an analogue bandwidth of \SI{350}{MHz} and a sampling rate of \SI{1.67}{GS/s}.

A total of 128 waveforms were averaged, and the baseline of \SI{-24.1}{mV}, measured without illumination, was subtracted. Assuming a single-exponential decay, the offset-corrected signal follows $\ln S(t)=\ln A-t/\tau$. The onset of the optical pulse occurred approximately \SI{30}{ns} after the electronic laser trigger. The approximately linear region from \SI{250}{ns} to \SI{600}{ns} after the trigger was fitted by linear least squares. This interval excluded the optical excitation pulse and the leading-edge response. The decay time was calculated from the fitted slope $m$ as $\tau=-1/m$.

\subsection{Synchrotron measurements}

The active beamstop was evaluated at the P14.EH2 (T-REXX) end station operated by EMBL Hamburg at the PETRA~III storage ring (DESY, Hamburg). The measurements were performed at an X-ray energy of \SI{12.7}{keV}. With the built-in collimator and aperture inserted, the beam diameter at the sample position was approximately \SI{10}{\micro m}, with an incident photon flux of approximately $1 \times 10^{12}\,\mathrm{ph/s}$. With the collimator and aperture retracted, the beam size was approximately \SI{10}{\micro m}~$\times$~\SI{30}{\micro m}, with a photon flux of approximately $3 \times 10^{12}\,\mathrm{ph/s}$.

The active beamstop was mounted on the standard end-station instrumentation (Arinax, Moirans, France) and positioned using its integrated actuators. For the initial alignment, a large-area photodiode was placed downstream of the beamstop, and the beamstop position was adjusted to maximize the blocking of the transmitted direct-beam signal. Fine alignment was subsequently performed by maximizing the output signal of the active beamstop.

The output signal from the amplifier PCB was transmitted through approximately \SI{15}{m} of RG-58 coaxial cable to a \SI{50}{\ohm}-terminated Keysight DSOX3034T oscilloscope in the control hutch. Acquisition settings, including record length and waveform averaging, were adjusted according to the storage-ring filling mode and are specified for the individual measurements below.

Measurements were performed during PETRA~III operation in the 40-bunch and 480-bunch filling modes. The storage-ring revolution period was \SI{7.68}{\micro s}, corresponding to nominal bunch separations of \SI{192}{ns} and \SI{16}{ns}, respectively. A timing signal synchronised with the storage-ring revolution frequency was used to trigger the oscilloscope.

\subsection{X-ray chopper measurements}

A mechanical X-ray chopper developed for an experiment aimed at reducing radiation damage to biological samples was operated at a rotation frequency of \SI{24}{Hz}. Its geometrically defined open fraction of 1/220 corresponded to an expected X-ray transmission window of approximately \SI{190}{\micro s} per chopper cycle. The active beamstop was positioned downstream of the experimental interaction point and intercepted the transmitted direct beam in its intended beamstop configuration. The beamstop signal and a \SI{100}{\micro s}-wide output pulse from the delay generator controlling the experimental sequence were recorded simultaneously using a Keysight DSOX3034T oscilloscope.

\section{Results and discussion}\label{sec:res_dis}

\subsection{Electronic noise and shielding}

The electronic noise was measured with and without the copper shielding installed around the photodiode and input-amplifier region. Without shielding, the standard deviation and peak-to-peak amplitude of the output voltage were \SI{4.21}{mV} and \SI{70}{mV}, respectively. With the shielding installed, the corresponding values decreased to \SI{3.74}{mV} and \SI{32}{mV}. The relatively small reduction in the standard deviation, compared with the more pronounced reduction in the peak-to-peak amplitude, indicates that the shielding primarily suppressed intermittent electromagnetic pickup rather than broadband electronic noise.

\subsection{Temporal response under pulsed-laser excitation}

\begin{figure}[thb]
    \centering
    \includegraphics[width=1.0\textwidth]{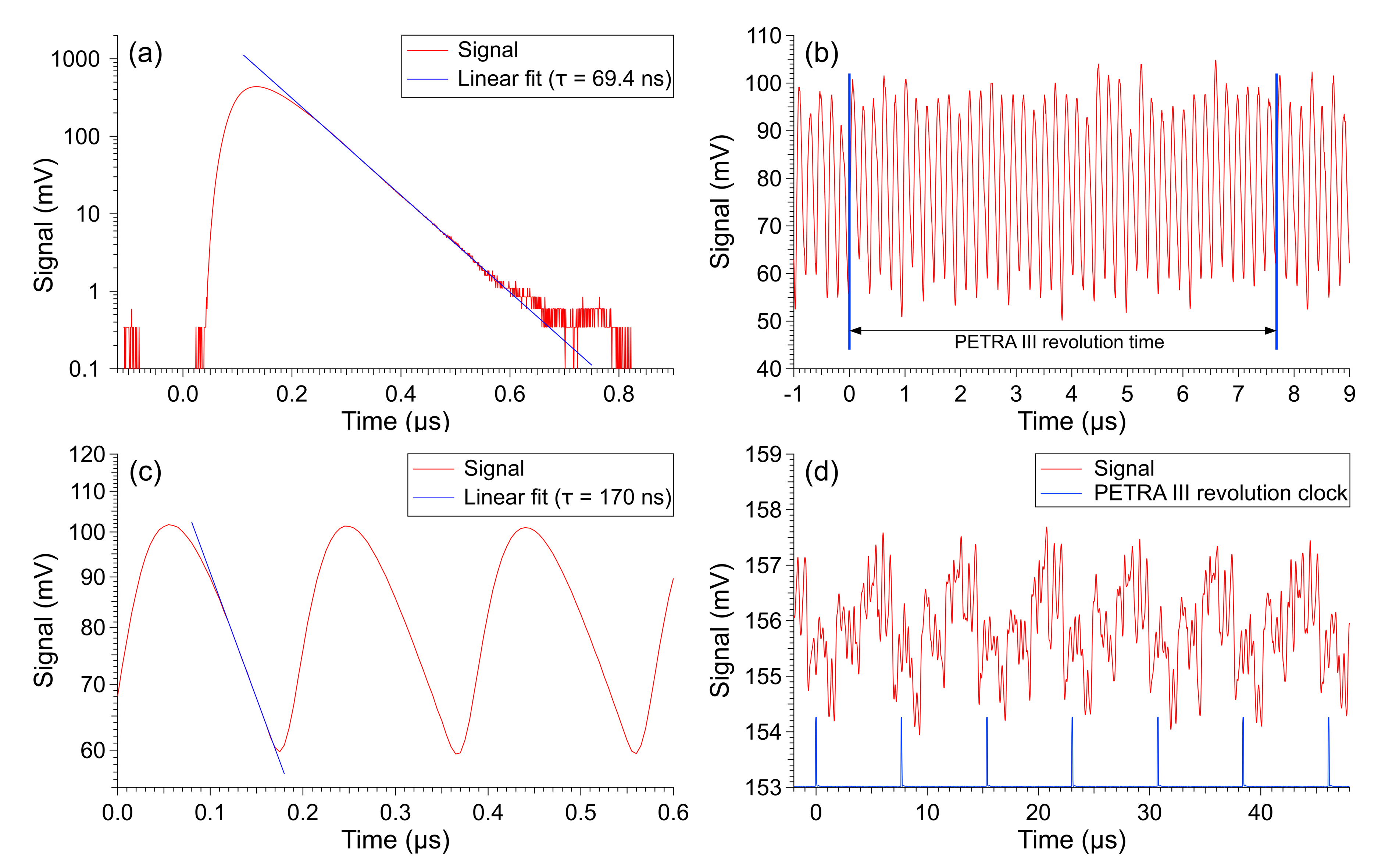}%
    \caption{Active beamstop output waveforms after subtraction of a common baseline offset.
    (a)~Response to pulsed-laser excitation of the Ce:YAG scintillator.
    (b)~Single-acquisition waveform recorded during PETRA III operation in 40-bunch mode, with the beam defined by the built-in collimator and aperture.
    (c)~Expanded view of the bunch response in 40-bunch mode averaged over 128 acquisitions and recorded with the same collimator and aperture configuration as in panel~(b).
    (d)~Signal recorded over approximately six storage-ring revolutions in 480-bunch mode and averaged over 64 acquisitions; the collimator and aperture were retracted for this measurement.}
    \label{fig:osc_waveforms}
\end{figure}

Figure~\ref{fig:osc_waveforms}a shows the response of the active beamstop to pulsed optical excitation of the Ce:YAG scintillator. The offset-corrected waveform exhibits a linear region on the semi-logarithmic scale, indicating that the selected falling-edge interval is dominated by a single exponential component. The fit described in Section~\ref{subsec:pulsed_laser_meas} yielded a decay time of \SI{69.4}{ns}, in close agreement with the commonly reported Ce:YAG luminescence decay time of approximately \SI{70}{ns} \cite{Blasse1967,Blasse1994,Bachmann2009}. The fitted component is therefore attributed to the luminescence decay of the scintillator, indicating that the readout chain was sufficiently fast not to appreciably limit the response within the selected fitting interval.

The earlier part of the waveform deviates from single-exponential behaviour. This deviation cannot be explained by the specified slew rate of the operational amplifier alone and may reflect a combination of \SI{450}{nm} excitation light coupled directly into the fibre, the finite response of the photodiode and amplifier, and possible transient saturation or overload recovery in the readout chain. As these contributions were not investigated independently, the fitting interval from \SI{250}{ns} to \SI{600}{ns} was selected to exclude the non-exponential initial response.

\subsection{Synchrotron beam-monitoring performance}

The active beamstop was tested during PETRA III operation in both the 40-bunch and 480-bunch filling modes. In 40-bunch mode, signals originating from individual electron bunches were clearly distinguishable in a single acquisition, as shown in Fig.~\ref{fig:osc_waveforms}b. The built-in collimator and aperture were inserted for this measurement. The typical peak amplitude was approximately \SI{80}{mV}, demonstrating that the \SI{192}{ns} bunch spacing could be detected without waveform averaging.

To examine the shape of the bunch response more closely, a waveform was subsequently acquired under the same beam-defining conditions by averaging 128 acquisitions [Fig.~\ref{fig:osc_waveforms}c]. Although the bunch separation was almost three times the decay time measured under pulsed-laser excitation, the signal did not return to its baseline before the arrival of the following bunch. A single-exponential fit to the selected interval between bunches yielded an apparent decay time of approximately \SI{170}{ns}, substantially longer than the value obtained under laser excitation.

This apparent time constant should not be interpreted directly as the intrinsic scintillation lifetime. The synchrotron waveform represents the periodically repeated response of the complete detection chain and may include overlapping contributions from preceding bunches, slower scintillation components, and amplifier or cable effects. Slow components associated with trapping and delayed radiative recombination have been reported for Ce:YAG and may also contribute to the observed response \cite{Nikl_2006}. Measurements using isolated X-ray pulses or a model accounting for the periodic excitation would be required to separate these contributions.

In 480-bunch mode, the nominal bunch separation of \SI{16}{ns} was substantially shorter than the temporal response of the active beamstop, and signals from individual bunches could not be distinguished. Figure~\ref{fig:osc_waveforms}d shows a waveform averaged over 64 acquisitions, recorded with the built-in collimator and aperture retracted. The signal exhibited a modulation synchronised with the storage-ring revolution period. The peak-to-peak modulation amplitude was approximately \SI{2.5}{\percent} of the mean signal. This behaviour is consistent with a slightly non-uniform filling pattern of the storage ring; however, a quantitative comparison was not possible because the bunch-current data available at the time were not recorded.

\subsection{In-situ monitoring of an X-ray chopper}

The active beamstop signal provided a direct temporal indication of X-ray transmission through the mechanical chopper. A representative oscilloscope recording is shown in Fig.~S5. The \SI{100}{\micro s}-wide delay-generator pulse triggering the actual data collection of the experiment was positioned within the approximately \SI{190}{\micro s} transmission window of the chopper as measured by the active beamstop. The signal was therefore used to adjust the timing of the experimental sequence relative to the chopper opening and remained available for monitoring during diffraction data acquisition.

The waveform was recorded at an earlier stage of the device development, before the final amplifier electronics and the shielding were implemented, and consequently exhibits a lower signal-to-noise ratio than the measurements reported above. The recording is intended to demonstrate the timing procedure rather than to provide a quantitative measurement of the transmission-window duration or timing jitter.

\section{Conclusion and outlook}

We have developed a fibre-coupled active beamstop for simultaneous interception and time-resolved monitoring of the direct X-ray beam in synchrotron diffraction experiments. The use of a \SI{700}{\micro m}-wide Ce:YAG scintillator at the beamstop tip, with optical-fibre transport to a remotely positioned photodiode and amplifier, minimises the dimensions of the detection element within the diffraction cone. The device provides an electrical signal correlated with the intensity of the transmitted direct beam and can therefore serve as a local downstream monitor of beam intensity and timing.

Under pulsed-laser excitation, the complete scintillator and readout system exhibited a falling-edge decay time of \SI{69.4}{ns}, consistent with the reported scintillation lifetime of Ce:YAG. At the P14.EH2 end station of PETRA~III, individual bunch signals separated by \SI{192}{ns} were distinguishable in 40-bunch mode without waveform averaging. In 480-bunch mode, the \SI{16}{ns} bunch spacing could not be resolved, but a modulation synchronised with the storage-ring revolution period remained observable. The active beamstop signal was also used to position a \SI{100}{\micro s}-wide delay-generator pulse within an approximately \SI{190}{\micro s} X-ray transmission window produced by a mechanical chopper. This demonstrates the practical use of the device for in-situ synchronization and monitoring during diffraction data acquisition.

Further work should include quantitative characterisation of the direct-beam attenuation, signal linearity with X-ray flux, spatial sensitivity across the scintillator and long-term radiation stability. The temporal response may be improved through the use of scintillators with shorter decay times and reduced afterglow, while optimization of the scintillator--fibre and fibre--photodiode interfaces could increase the collected light without enlarging the beamstop tip. These developments could extend the applicable range of bunch structures and improve sensitivity at lower photon fluxes.

%
    

\begin{acknowledgements}

The authors would like to thank G.~Bourenkov and M.~Agthe (EMBL Hamburg) and J.-P.~Leimkohl and H.~Schikora (Max Planck Institute for the Structure and Dynamics of Matter) for their continuous support and helpful discussions during the implementation and improvement of this instrument.
The authors also thank M.~A.~Klureza (University of Hamburg) for providing the diffraction image.
All X-ray experiments were conducted on the T-REXX endstation of the EMBL beamline P14 at the PETRA III synchrotron at DESY, Hamburg (https://www.embl-hamburg.de/services/mx/).

FT and DvS designed and performed the experiment. FT and MK designed the amplifier electronics and analysed the data. FT conceptualized the active beamstop and wrote the manuscript. All authors discussed the results and revised the manuscript.

\end{acknowledgements}

\begin{funding}
This research received no specific grant from any funding agency in the
public, commercial or not-for-profit sectors.
\end{funding}

\ConflictsOfInterest{The authors declare that there are no conflicts of interest.}

\DataAvailability{The data supporting the findings of this study are available
within the article and its supporting information. Additional data are available
from the corresponding authors upon request.}

\bibliography{iucr} 

\appendix
\setcounter{figure}{0}
\setcounter{table}{0}
\section*{\underline{Supporting Information for:}\\
A compact fibre-coupled active beamstop for fast time-resolved monitoring of synchrotron X-ray beams}



\renewcommand{\thefigure}{S\arabic{figure}}

\begin{figure}[h]
\centering
\includegraphics[width=1.0\linewidth]{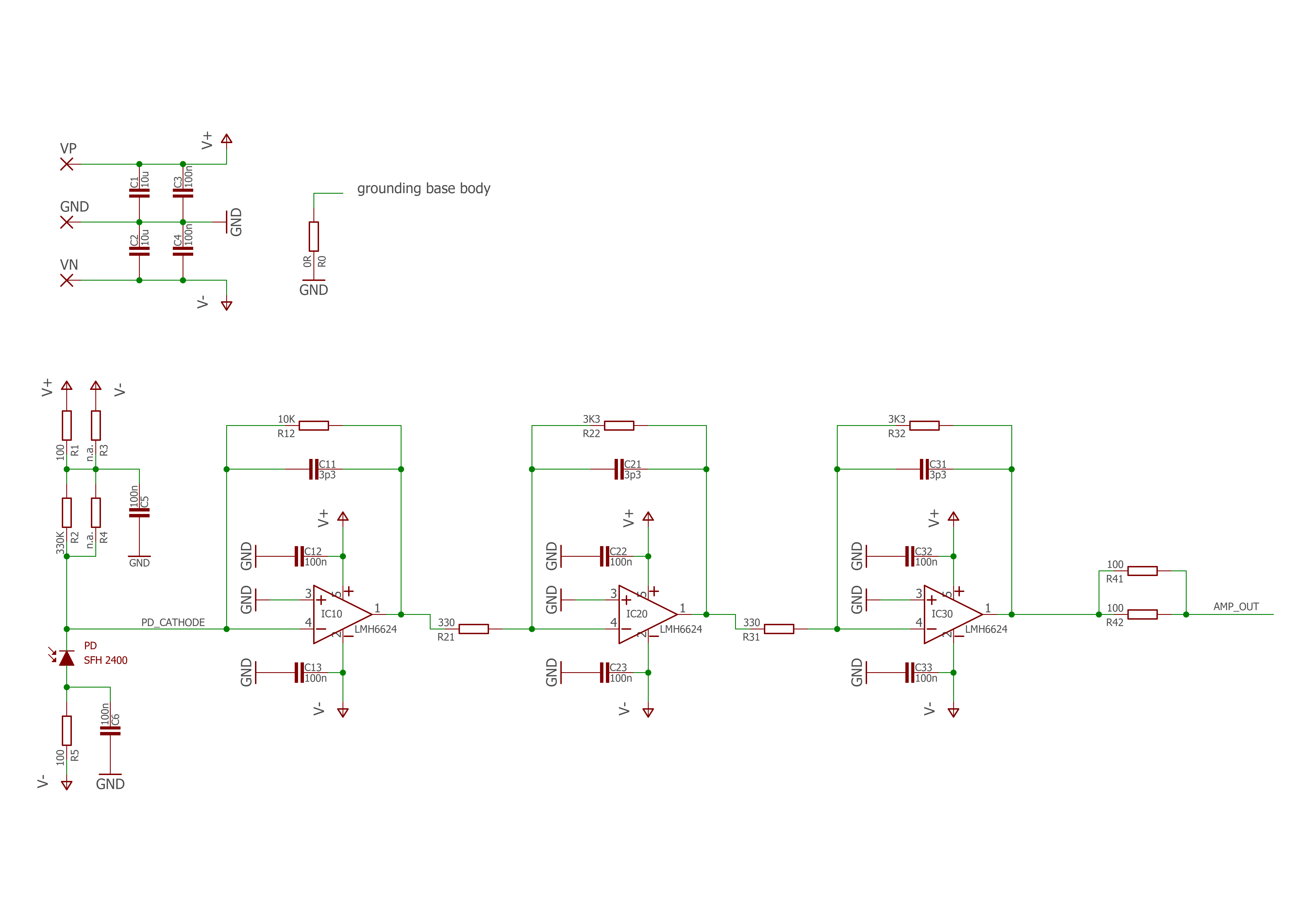}
\caption{Schematic diagram of the photodiode and amplifier electronics used for the measurements reported in this work. The first operational-amplifier stage is configured as a transimpedance amplifier, followed by two voltage-amplification stages. Resistors $R_1$--$R_4$ provide optional offset compensation. The component values and configuration shown correspond to those used in the reported measurements.}
\label{SI:schematics}
\end{figure}

\begin{figure}[h]
\centering
\includegraphics[width=1.0\linewidth]{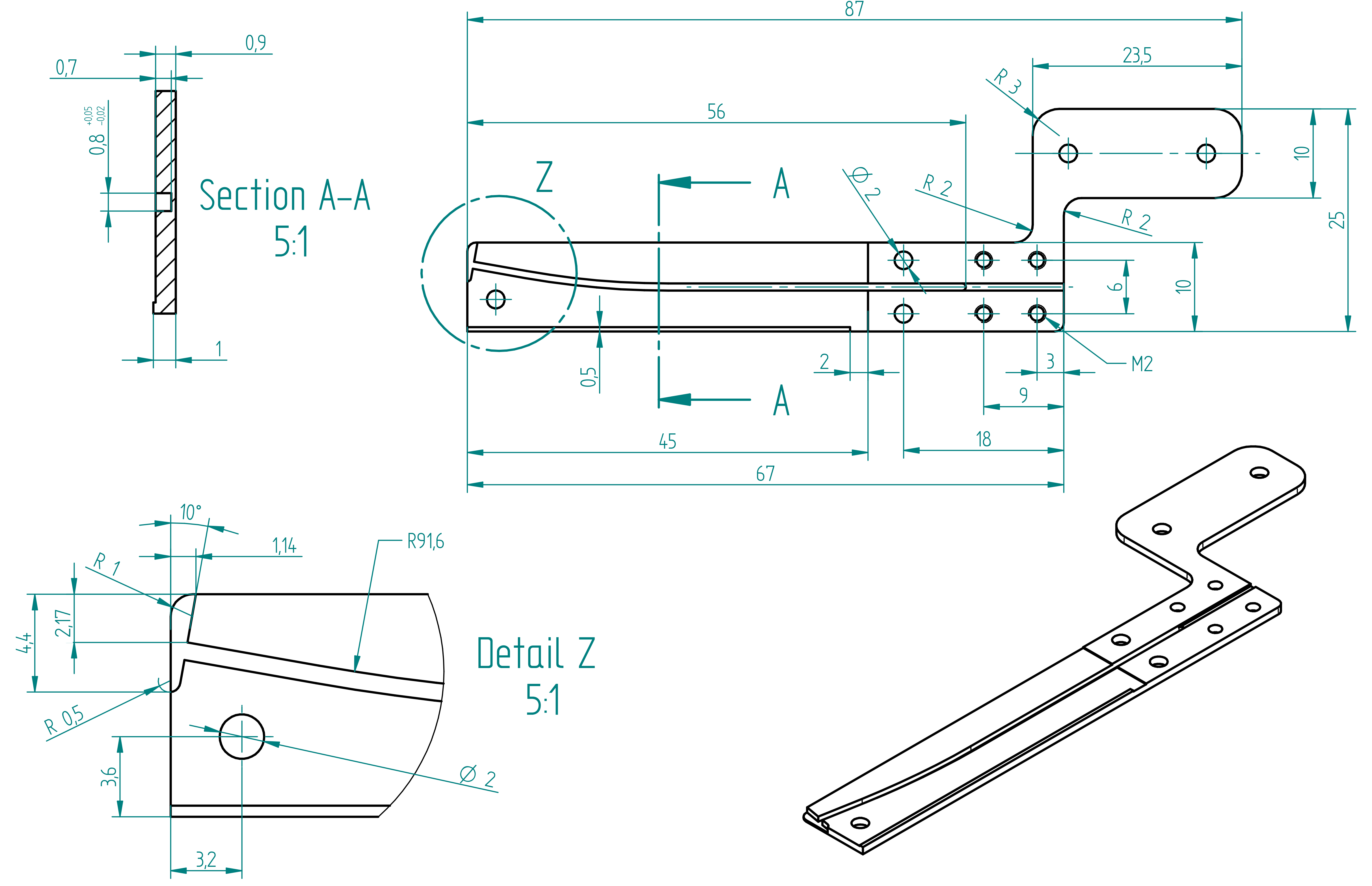}
\caption{Technical drawing of the beam-stop blade. The holes with a diameter of \SI{2}{mm} were included solely to facilitate manufacturing and have no function in the assembled device. Dimensions are given in millimetres.}
\label{SI:drawing}
\end{figure}

\begin{figure}[h]
\centering
\includegraphics[width=0.45\linewidth]{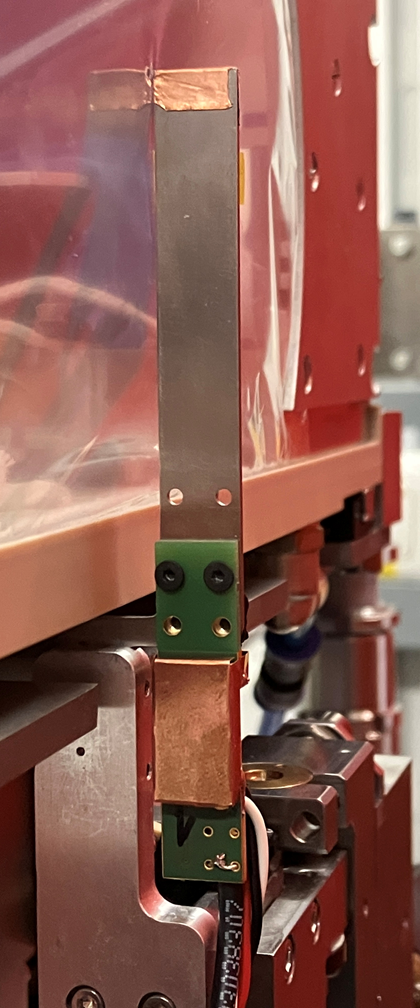}
\hspace{5mm}
\includegraphics[width=0.45\linewidth]{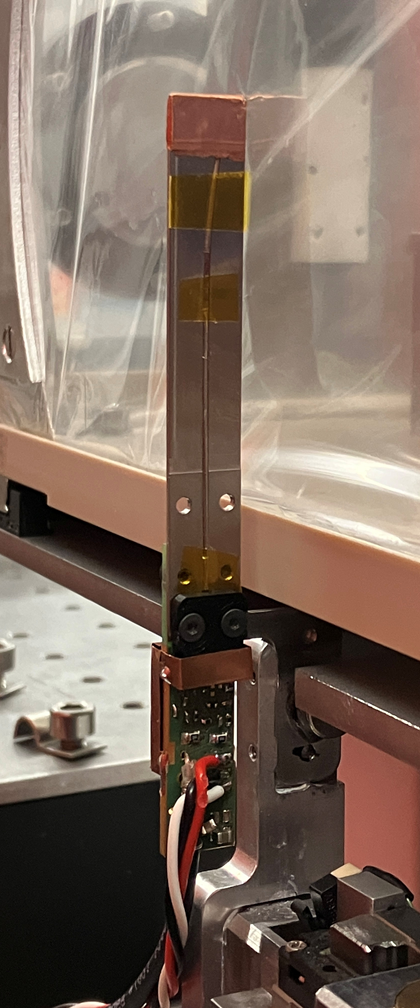}
\caption{Photographs of the active beam stop installed at the P14.EH2 end station of PETRA~III: (left)~back view and (right)~front view. In the left panel, the incident X-ray beam propagates from left to right.}

\label{SI:img_setup}
\end{figure}

\begin{figure}[h]
\centering
\includegraphics[width=1.0\linewidth]{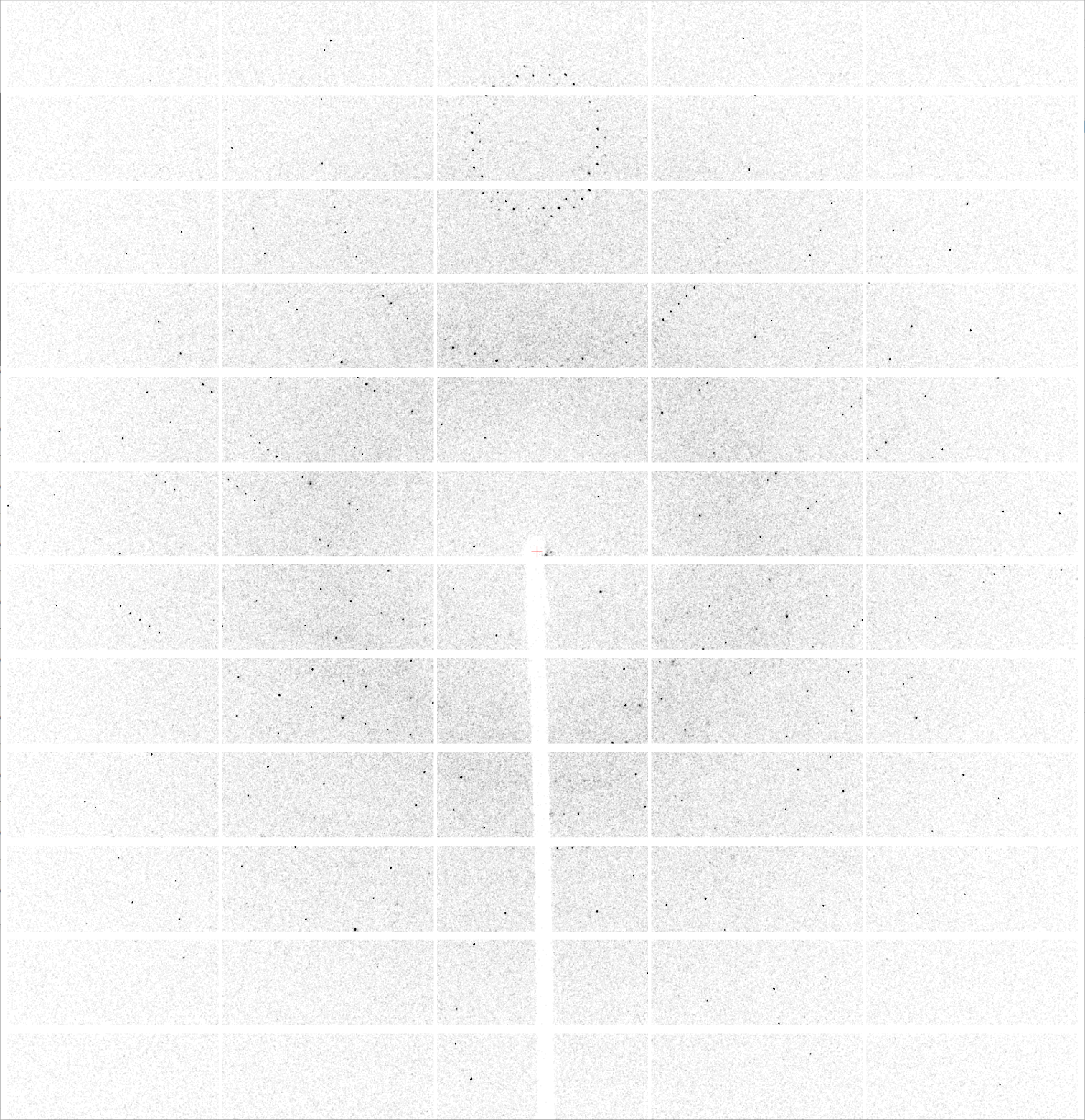}
\caption{Representative diffraction image from a hen egg-white lysozyme crystal acquired with the active beam stop installed downstream of the sample. The image demonstrates operation of the device in its intended beam-stop configuration during diffraction data acquisition. The X-ray energy was \SI{12.7}{keV}.}
\label{SI:img_detector}
\end{figure}

\begin{figure}[h]
\centering
\includegraphics[width=1.0\linewidth]{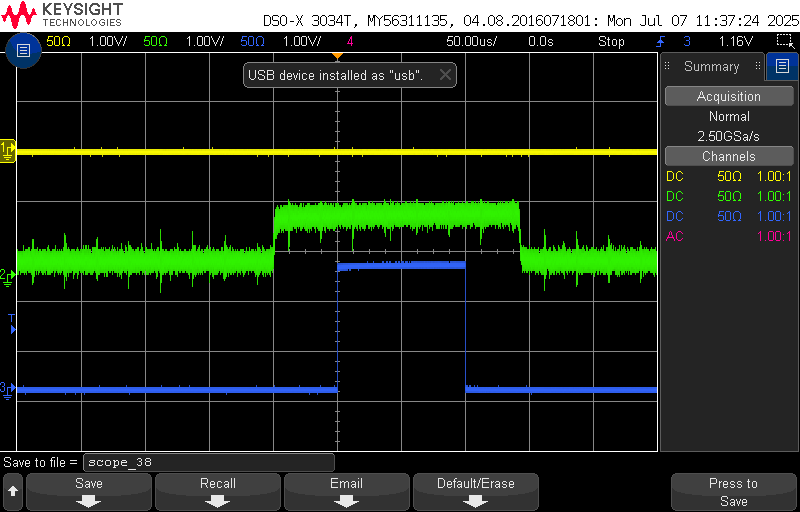}
\caption{Oscilloscope screenshot recorded during synchronization of the experimental sequence with the mechanical X-ray chopper. Trace~2 (green) shows the active beam-stop signal and indicates an X-ray transmission window of approximately \SI{190}{\micro s}. Trace~3 (blue) shows the \SI{100}{\micro s}-wide delay-generator signal, which was visually centred within the X-ray transmission window. The screenshot was acquired during an earlier stage of device development and is presented as a qualitative demonstration of the synchronization procedure. The numerical waveform data were not retained for further quantitative analysis.}
\label{SI:Xray_chopper}
\end{figure}

\end{document}